\documentclass[english,preprint,prd]{revtex4}
\usepackage[T1]{fontenc}
\usepackage[latin1]{inputenc}
\usepackage{amsmath}
\usepackage{graphicx}
\usepackage{amssymb}
\usepackage{dcolumn}
\usepackage{bm}
\usepackage{amsfonts}
\usepackage{babel}

\begin{document}
\title{Production of Doubly Charged Higgs Bosons
at Linear $e^{-}e^{-}$
Colliders}

\author{Orhan \c{C}ak{\i}r\thanks{e-mail:
ocakir@science.ankara.edu.tr}}

\address{Ankara University, Faculty of Sciences,
Department of Physics, 06100,
Tandogan, Ankara, Turkey. }

\begin{abstract}
Production of doubly charged Higgs bosons via the s-channel process
$e^{-}e^{-}\rightarrow H^{--}\rightarrow l^{-}l^{-}$ at the future
linear collider energies is studied by taking into account initial
state radiation and beamsstrahlung (ISR+BS), final state radiation
(FSR) and detector smearing effects. The discovery bounds of lepton
flavour conserving and violating couplings are obtained for doubly
charged Higgs bosons. It is found that the future linear colliders
with the center of mass energies $\sqrt{s}=500$ GeV and
$\sqrt{s}=3000$ GeV will be able to probe the doubly charged Higgs
bosons with diagonal couplings down to $10^{-4}$ and $10^{-3}$,
respectively.
\end{abstract}



\maketitle

\section{Introduction}

There is a consensus in the particle physics community that the
next large scale project should be a linear collider. The
International Linear Collider (ILC) \cite{1} is an $e^+e^-$ linear
collider with a center of mass energy of $\sqrt{s}=500$ GeV and
the luminosity $L\approx 10^{34}$cm$^{-2}$s$^{-1}$. Beyond this
baseline machine, upgrades and options are envisaged whose weight,
priority and realization will depend upon the results obtained at
the Large Hadron Collider (LHC) and the baseline ILC.

The physics results obtained in the first few years of running with
ILC, together with the results from the LHC will then define the
schedule for upgrades or other modes of operations (options). The
expected shutdowns to install the upgrades or options will not take
more than two years after an initial physics running time of at
least four years, including the commissioning of the upgrades or
options \cite{1}. The ILC can operate for part of its running time
in the electron mode where certain signals can be remarkably free
from backgrounds. The $e^-e^-$ running mode can also offer the high
polarizations (at least 80\%) for both initial states. This mode is
a particularly interesting feature of a high energy linear collider.
The main feature of this option is that the $e^{-}e^{-}$ initial
state is doubly charged and caries double lepton number. In this
mode, the standard model (SM) activity can be highly suppressed, and
clean signals from any non-standard physics can be searched. One of
the many promising processes which can be studied in this mode is
the resonant production of doubly charged scalars. Here we should
note that s-channel resonance production of doubly charged Higgs
bosons at $e^-e^-$ colliders dominates over their t-channel indirect
productions at $e^+e^-$ colliders. Therefore, an $e^{-}e^{-}$
collider could be used as a factory for doubly charged Higgs bosons,
where their masses, total decay widths and couplings could be
measured precisely. An additional feature often associated with
doubly charged Higgs boson is the possibility of lepton number
violation.

In many cases detailed measurements at a higher energy $e^+e^-$
collider would be needed to complement previous exploratory
observations. We think that further progress in particle physics
will necessitate clean experiments at multi-TeV energies which is
possible at $e^+e^-$ Compact Linear Collider (CLIC) \cite{2}. It is
anticipated that the center of mass energy of the CLIC as high as 5
TeV might eventually prove feasible. Thus, the pair production of
new particles could be detected in the $e^+e^-$ mode of operation up
to the mass of $\sqrt{s}/2$.

The Standard Model (SM) gives a good description of the known
fundamental particles, using $SU(3)_{C}\times SU(2)_{L}\times
U(1)_{Y}$ gauge group to describe their color and electroweak
interactions. The $SU(2)_{L}\times U(1)_{Y}$ electroweak gauge
symmetry is broken to $U(1)_{em}$ by the Higgs mechanism, but a
Higgs boson has yet to be observed \cite{3}. This is one of the
good reasons that other symmetry breaking mechanisms and extended
Higgs sectors have not been excluded from the theoretical point of
view. Some theories beyond the SM predict the existence of doubly
charged Higgs bosons, including in left-right symmetric models
\cite{4}, Higgs triplet models \cite{5}, and little Higgs models
\cite{6}. In the compositeness models they can be considered as
the doubly charged scalar bileptons \cite{7,7-1}. The left-right
symmetric (LR) electroweak theory based on the gauge symmetry
group $SU(2)_{L}\times SU(2)_{R}\times U(1)_{B-L}$ was proposed to
offer a dynamical solution to the parity violation of weak
interactions. The presence of triplet representations of Higgs
fields, i.e. $H_{R}$ triplet of $SU(2)_{R}$ and $H_{L}$ triplet of
$SU(2)_{L}$, provides a simple explanation to the lightness of
neutrinos via the see-saw mechanism \cite{8}, consistent with
recent data on neutrino oscillations \cite{9}. In addition,
indication for neutrino oscillations necessarily violate lepton
flavor symmetry \cite{10}. It has been emphasized that the see-saw
mechanism could lead to doubly charged Higgs bosons with masses
accessible to current and future colliders.

Direct searches for doubly charged Higgs bosons have been
performed by the OPAL \cite{11}, L3 \cite{12} and DELPHI \cite{13}
Collaborations at LEP, and they have excluded $H^{\pm\pm}$ bosons
below masses of about 100 GeV, assuming exclusive $H^{\pm\pm}$
decay to a given dilepton channel. Recent searches by the CDF and
D0 Collaborations at the Fermilab Tevatron in the $\mu\mu$ channel
have excluded $H_{L}^{\pm\pm}$($H_{R}^{\pm\pm}$) below a mass of
136 (113) GeV \cite{14} and 118 GeV \cite{15} at the 95\%
confidence level (C.L.), respectively. There are unknown
parameters on which the obtained constraints depend: the mass of
the doubly charged Higgs boson $m_{H^{\pm\pm}}$ and the coupling
constants $h_{ij}$, where $i,j=e,\mu,\tau$. The $H^{\pm\pm}$
couplings $h_{ij}$ to electrons and muons are experimentally
constrained by the absence of $H^{\pm\pm}$ production in
$e^{+}e^{-}$ collisions and nonobservation of the decay
$\mu\rightarrow3e$
($h_{ee}h_{e\mu}<3.2\times10^{-11}m_{H^{\pm\pm}}^{2}$ GeV$^{-2}$)
and decay $\mu\rightarrow e\gamma$
($h_{\mu\mu}h_{e\mu}<2\times10^{-10}m_{H^{\pm\pm}}^{2}$
GeV$^{-2}$) \cite{16}. From the Bhabba scattering the relevant
bound is given as $h_{ee}^{2}<6.0\times10^{-6}m_{H^{\pm\pm}}^{2}$
GeV$^{-2}$ \cite{17} and from $(g-2)_{\mu}$ measurements it is
$h_{\mu\mu}^{2}<2.5\times10^{-5}m_{H^{\pm\pm}}^{2}$ GeV$^{-2}$
\cite{18}. More stringent bounds exist from the
muonium-antimuonium transition \cite{19} in the form of the
product of the couplings,
$h_{ee}h_{\mu\mu}<2.0\times10^{-7}m_{H^{\pm\pm}}^{2}$ GeV$^{-2}$.
Similar bounds are given for both lepton flavor conserving
couplings $h_{ee}^{2}<3.5\times10^{-5}m_{H^{--}}^{2}$ GeV$^{-2}$
\cite{19-1} and violating couplings
$h_{e(\mu,\tau)}^{2}<1\times10^{-6}m_{H^{--}}^{2}$ GeV$^{-2}$
\cite{19-2} of doubly charged scalar bileptons (as doubly charged
Higgs boson $H^{--}$) from Bhabba scattering with LEP data. The
experimental constraints on the couplings weaken with increasing
$H^{\pm\pm}$ mass.

It is reasonable to assume that below the scale of 500 GeV the
left-right symmetry is broken \cite{19_2_1} due to the strength of
the left/right handed interactions and the form of CKM matrix of the
right-handed interactions. It is entirely reasonable to suppose that
a $H^{--}$ in the $m_{H^{--}}<500$ GeV mass range relevant for a
$\sqrt{s}=500$ GeV $e^-e^-$ collider would already have been
observed at the LHC, regardless of the magnitude of the coupling
$h_{ll}$.

In $e^{-}e^{-}$ collisions the production of doubly charged Higgs
bosons via lepton number conserving and violating processes have
been studied in \cite{7-1,19-3}. A work for the usefulness of
looking at associated monoenergetic photon as signals for a doubly
charged scalars at a linear $e^-e^-$ collider with $\sqrt{s}=1$
TeV has been presented in \cite{19-4}. Recently, doubly charged
scalar bileptons (as doubly charged Higgs bosons) have been
studied via $e^{-}e^{-}\rightarrow e^{-}e^{-},\mu^{-}\mu^{-}$
processes with a resonance scan of the collider energies up to 2
TeV \cite{19-5}.

In this paper, we study in detail the production of doubly charged
Higgs bosons $H_{L,R}^{--}$ through s-channel in $e^{-}e^{-}$
collision including initial state radiation (ISR)+beamstrahlung
(BS) and final state radiation (FSR). We also include the smearing
effects of a detector on the four-momentum of final state leptons.
We analyze the signature for identifying an isolated same sign,
planar and $p_{T}$ balanced two-lepton events. We also calculate
the discovery bounds of couplings and masses for doubly charged
Higgs bosons with lepton flavour conserving and violating
couplings.

\section{Doubly Charged Leptonic Interactions}

The doubly charged Higgs bosons have the following Yukawa couplings
to the leptons:

\begin{equation}
L_{Y}=h_{R,ij}l_{iR}^{T}C\sigma_{2}H_{R}l_{j,R}+h_{L,ij}l_{iL}^{T}C\sigma_{2}H_{L}l_{j,L}+h.c.\label{eq:1}\end{equation}
where $l_{i,L(R)}^{T}=(\nu_{i},l_{i})_{L(R)}$ are two-component left
or right handed lepton fields with flavour index $i$; C is the
charge conjugation matrix; $\sigma_{2}$ is the Pauli matrix. $H_{L}$
and $H_{R}$ are the $SU(2)_{L}$ and $SU(2)_{R}$ triplets,
respectively. The Higgs triplets

\begin{equation}
H_{L,R}=\left(\begin{array}{cc}
H_{L,R}^{-}/\sqrt{2} & H_{L,R}^{--}\\
H_{L,R}^{0} & -H_{L,R}^{-}/\sqrt{2}\end{array}\right)\label{eq:2}\end{equation}
acquire non-vanishing vacuum expectation values (vev) given by

\begin{equation}
<H_{L,R}>=\frac{1}{\sqrt{2}}\left(\begin{array}{cc}
0 & 0\\
v_{L,R} & 0\end{array}\right)\label{eq:3}\end{equation}
where the left-handed triplet vev $v_{L}$ is kept to be small due
to its contribution to the $\rho$-parameter, $\rho\simeq(1+2v_{L}^{2}/\kappa^{2})/(1+4v_{L}^{2}/\kappa^{2})$
where $\kappa$ is related to the vev of bidoublet scalar field. The
vev $v_{R}$ gives masses to $W_{R}$ as well as doubly charged scalars
$H_{L,R}^{--}$. The strength of the interaction is scaled by the
symmetric unknown constants $h_{ij}$ which, in general, are not flavour
diagonal allowing for lepton number violating interactions. Therefore
there are six possible couplings:

\begin{equation}
h=\left(\begin{array}{ccc}
h_{ee} & h_{e\mu} & h_{e\tau}\\
h_{\mu e} & h_{\mu\mu} & h_{\mu\tau}\\
h_{\tau e} & h_{\tau\mu} & h_{\tau\tau}\end{array}\right)\label{eq:4}\end{equation}

Using the Lagrangian (1) with the assumption on the structure of
the leptonic couplings, the leptonic decay widths for the doubly
charged Higgs bosons are given by \begin{equation}
\Gamma_{H^{--}}=\frac{m_{H^{--}}}{8\pi}\underset{i,j}{\sum}h_{ij}^{2}\label{eq:5}\end{equation}
We calculate the leptonic decay widths for $H^{--}$ as
$\Gamma_{H^{--}}=0.597$ GeV taking the mass $m_{H^{--}}=500$ GeV
and assuming universal flavour diagonal couplings
$h_{ij}=0.1\delta_{ij}.$ Since we consider only leptonic decay
channels of doubly charged Higgs bosons, calculated decay widths
become narrower for smaller couplings and masses. Other decay
channels including gauge bosons can be neglected assuming the vev
of the neutral member of the triplet is small and taking into
account the limits on the model's couplings and the mass spectrum
of the Higgs bosons. The two doubly charged Higgs bosons
$H_{L,R}^{--}$ can have different chiral couplings to leptons.
Their masses are expected to be comparable with each other because
they derive both from similar terms of scalar potential. Since
their production processes are the same we concentrate on the
production of $H^{--}$($H_{L}^{--}$ or $H_{R}^{--}$).

\section{Production Cross Sections}

\begin{figure}
\begin{center}\includegraphics[
  width=14cm,
  height=1.6cm]{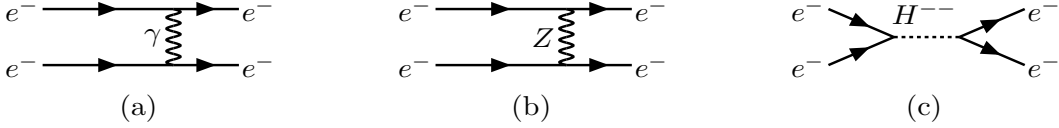}\end{center}

\caption{Feynman diagrams for the process $e^{-}e^{-}\rightarrow e^{-}e^{-}$
including doubly charged Higgs boson $H^{--}$ exchange. \label{fig1}}
\end{figure}

For the process $e^{-}e^{-}\rightarrow e^{-}e^{-}$ including doubly
charged Higgs bosons $H_{L,R}^{--}$, as depicted in Fig. \ref{fig1},
the corresponding differential cross section can be obtained as

\begin{eqnarray}
\frac{d\sigma}{dt} & = & \frac{1}{8\pi}\left\{ F_{LL}\left|\frac{g_{e}^{2}s}{t(s+t)}+\frac{g_{e}^{2}C_{LL}(s+2m_{Z}^{2})}{(t-m_{Z}^{2})(s+t+m_{Z}^{2})}+\frac{h_{L,ee}^{2}}{(s-m_{H_{L}^{--}}^{2}+im_{H_{L}^{--}}\Gamma_{H_{L}^{--}})}\right|^{2}\right.\nonumber \\
 &  & +F_{RR}\left|\frac{g_{e}^{2}s}{t(s+t)}+\frac{g_{e}^{2}C_{RR}(s+2m_{Z}^{2})}{(t-m_{Z}^{2})(s+t+m_{Z}^{2})}+\frac{h_{R,ee}^{2}}{(s-m_{H_{R}^{--}}^{2}+im_{H_{R}^{--}}\Gamma_{H_{R}^{--}})}\right|^{2}\nonumber \\
 &  & \left.+g_{e}^{4}F_{RL}\left[\frac{t}{s+t}-\frac{C_{RL}t}{(s+t+m_{Z}^{2})}\right]^{2}+g_{e}^{4}F_{LR}\left[\frac{s+t}{t}+\frac{C_{LR}(s+t)}{(t-m_{Z}^{2})}\right]^{2}\right\} \label{eq:6}\end{eqnarray}
where $\Gamma_{H_{L,R}^{--}}$ is the total decay width of
$H_{L,R}^{--}$ boson. The polarization factors $F_{RR,LL}=(1$$\pm
P_{1})(1\pm P_{2})/4$ and $F_{RL,LR}=(1-$$P_{1}P_{2})/2$ with the
polarizations $P_{1}$ and $P_{2}$ of the incoming electron beams.
The mixing angle dependent constants $C_{LL}=s_{w}^{2}/c_{w}^{2}$,
$C_{RR}=(1-2s_{w}^{2})^{2}/(2s_{w}c_{w})^{2}$ and
$C_{LR}=C_{RL}=(1-s_{w}^{2})/2c_{w}^{2}$. The abbreviations $c_{w}$
and $s_{w}$ are used for $\textrm{cos}\theta_{W}$ and
$\textrm{sin}\theta_{W}$, respectively. In Eq. (\ref{eq:6}), $g_e$
is the electromagnetic coupling constant; $s$ and $t$ are the
Mandelstam variables, and $m_{Z}$ is the $Z$ boson mass.

Here we assume that electron beam is $\sim 100\%$ left or right
polarized. This is a simplification which is motivated by the very
high polarization rate $80-90\%$ achievable at the $e^-e^-$ options
of the ILC and CLIC environment. The impact of an exact amount of
80\% left polarization results in the mixture of partial LL, LR/RL
and RR contributions (being LL 80\% dominant) to the cross section
given in (6).

For the processes $e^{-}e^{-}\rightarrow\mu^{-}\mu^{-}$ and
$e^{-}e^{-}\rightarrow\tau^{-}\tau^{-}$ only flavour diagonal
couplings contribute to the cross section. The resonance cross
section for the signal processes $e^{-}e^{-}\rightarrow
H^{--}\rightarrow l^{-}l^{-}$ (where $l^{-}$=$\mu^{-}$ or
$\tau^{-}$) can be obtained as
\begin{equation}
\sigma=\frac{8\pi\Gamma_{ee}\Gamma_{ll}}{m_{H^{--}}^2}\frac{s}{[(s-m_{H^{--}}^{2})^{2}+m_{H^{--}}^{2}\Gamma_{H^{--}}^{2}]}
\label{eq:7}\end{equation} if the decay width of doubly charged
Higgs boson is significantly larger than the beam energy spread
(typically $10^{-2}\sqrt{s}$). It is a good approximation to assume
monochromatic electron beams. However, in the narrow width
approximation ($\Gamma_{H^{--}}/m_{H^{--}}<0.01$), the cross section
formula (7) becomes
\begin{equation}
\sigma=\frac{8\pi^2}{m_{H^{--}}}B_{ee}B_{ll}\Gamma_{H^{--}}\delta(s-m_{H^{--}}^2)
\label{eq:8}\end{equation} depending on the total decay width and
branchings $B_{ll}$ of the doubly charged Higgs boson. Here, we
consider only leptonic decay channels of the $H^{--}$ and assume
that the energy spectrum of the luminosity is broader than the
narrow resonant width. It should be noted that the actual total
cross section can be obtained by convoluting $\sigma(\hat s)$ with
the initial and final state radiation functions, and the smearing
function.

We have implemented the doubly charged Higgs boson interaction
vertices into the Monte Carlo simulation program CompHEP \cite{20}
for decay width and cross section calculations. Since the center
of mass energy is not well defined due to initial state radiation,
beamstrahlung and the energy spread of the particles in the beam,
these effects modify the energy dependence of the resonance cross
section significantly. If the luminosity spectrum is well measured
the physical parameters can still be extracted from the resonance
scan. The scan can be done measuring the cross section at
different center of mass energies around the resonance point.

\begin{figure}
\begin{center}\includegraphics{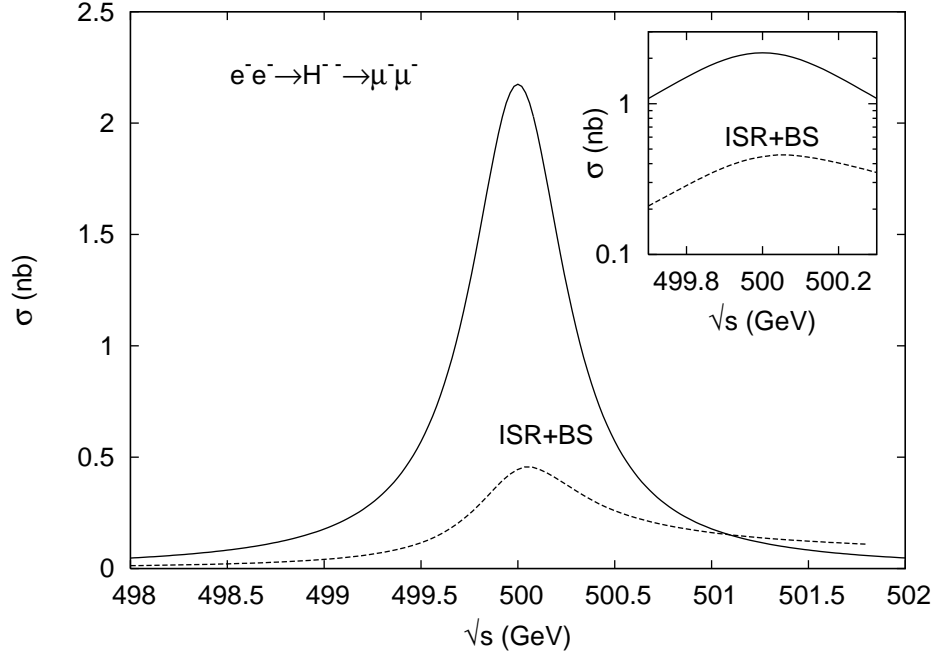}\end{center}

\caption{The production cross section for the process $e^{-}e^{-}\rightarrow H^{--}\rightarrow\mu^{-}\mu^{-}$
depending on the centre of mass energy of the $e^{-}e^{-}$ collider.
Solid line shows the cross section without the ISR+BS, dashed line
represent ISR+BS effect for the flavour diagonal couplings $h_{ee,\mu\mu}=0.1$\label{fig2}}
\end{figure}

\begin{figure}
\begin{center}\includegraphics{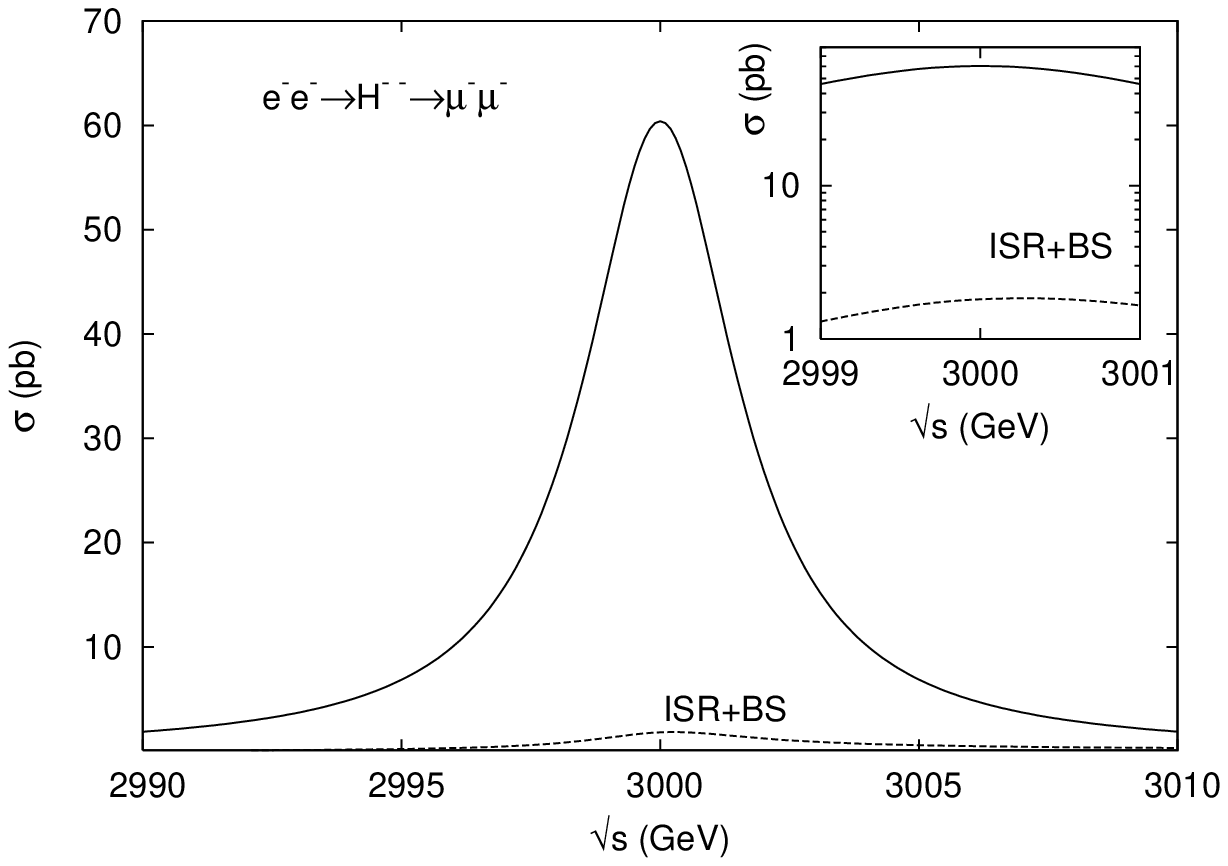}\end{center}

\caption{The cross section $\sigma(e^{-}e^{-}\rightarrow H^{--}\rightarrow\mu^{-}\mu^{-})$
depending on the center of mass energy around 3000 GeV. Solid line
for the cross section without the ISR+BS, dashed line represent ISR+BS
effect for $h_{ee,\mu\mu}=0.1$. \label{fig3}}
\end{figure}

In the case of the process $e^{-}e^{-}\rightarrow
H^{--}\rightarrow\mu^{-}\mu^{-}$ the cross section with and without
ISR+BS are calculated around the resonance point as shown in Fig.
\ref{fig2} and \ref{fig3}. The resonance peak shifts to the right (a
higher energy) and reduce by a factor of about 5 due to ISR+BS
effects. At the ILC based $e^{-}e^{-}$ collider with the center of
mass energy $\sqrt{s}=500$ GeV we obtain the resonance cross
sections: $\sigma=2.17\times10^{3}$ pb,
$\sigma_{ISR}=1.03\times10^{3}$ pb and
$\sigma_{ISR+BS}=4.45\times10^{2}$ pb; without radiation, with ISR
and with ISR+BS effects, respectively. For CLIC based $e^{-}e^{-}$
collider with the center of mass energy $\sqrt{s}=3000$ GeV this
effect become apparent from Fig. \ref{fig3}, as we find the signal
cross section $\sigma=60.41$ pb without radiation effects. In this
case ISR+BS affects the cross sections significantly as
$\sigma_{ISR+BS}=1.84$ pb. When calculating these effects we take
into account the beam parameters of the ILC and CLIC based
$e^{-}e^{-}$ colliders as shown in Table \ref{table1}.

\begin{table}

\caption{The beam parameters of ILC and CLIC based $e^{-}e^{-}$ colliders
used in the calculations of ISR and beamstrahlung. $N$ is the number
of particles in the electron bunch; $\sigma_{x,y,z}$ are the average
sizes of the electron bunches; $\Upsilon$ is the beamstrahlung parameter
and $N_{\gamma}$ is the average number of photons per electron. \label{table1}}

\begin{center}\begin{tabular}{lllll}
\hline
Collider &
&
 ILC ($e^{-}e^{-}$) &
&
 CLIC ($e^{-}e^{-}$)\tabularnewline
Parameters &
&
 500 GeV &
&
 3000 GeV\tabularnewline
\hline
$N(10^{10})$&
&
 2 &
&
 0.4\tabularnewline
$\sigma_{x+y}$ (nm) &
&
 661 &
&
 44\tabularnewline
$\sigma_{z}$ (mm) &
&
 0.3 &
&
 0.03\tabularnewline
$\Upsilon$&
&
 0.04 &
&
 8.07\tabularnewline
$N_{\gamma}$&
&
 1.22 &
&
 1.74  \tabularnewline
\hline
\end{tabular}\end{center}
\end{table}

\begin{figure}
\begin{center}\includegraphics{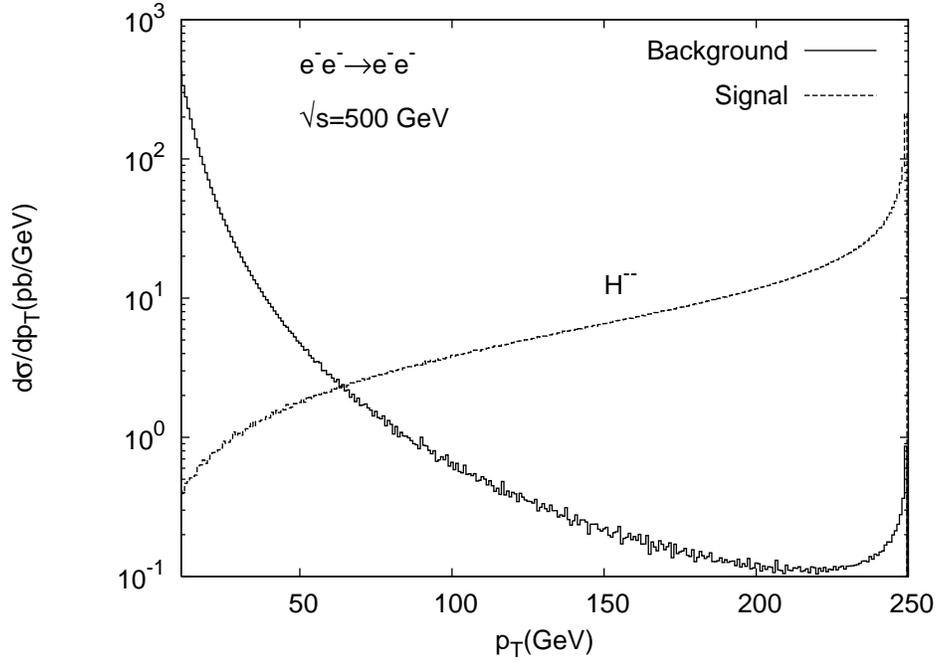}\end{center}

\caption{Transverse momentum distribution of final state electrons for the
signal ($h_{ee}=0.1$) and background processes at $\sqrt{s}=500$
GeV.\label{fig4}}
\end{figure}

\begin{figure}
\begin{center}\includegraphics{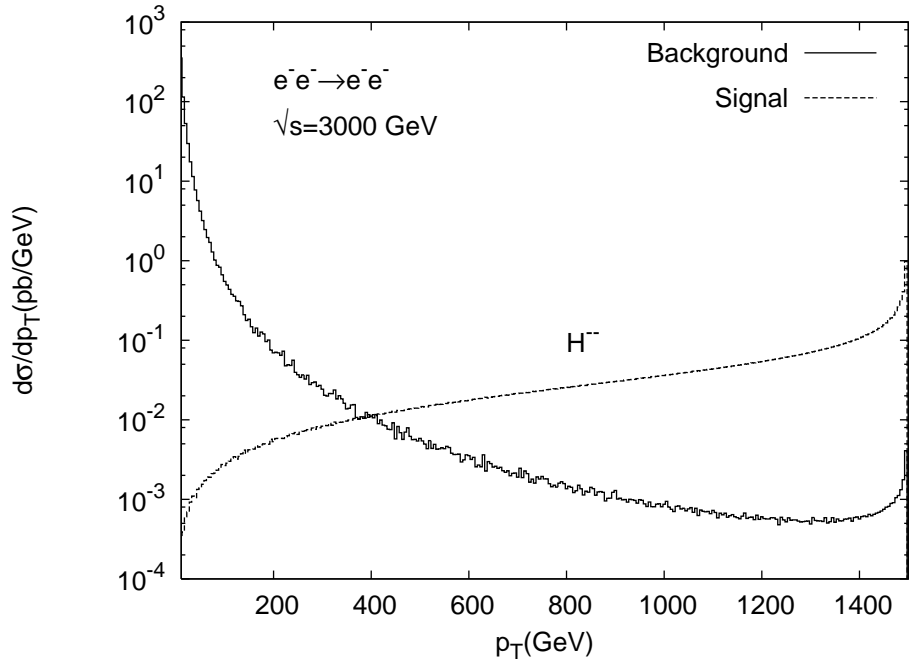}\end{center}

\caption{Transverse momentum distribution of final state electrons for the
signal ($h_{ee}=0.1$) and background processes at $\sqrt{s}=3000$
GeV.\label{fig5}}
\end{figure}

In Fig. \ref{fig4} and \ref{fig5}, we present the differential cross
sections ($d\sigma/dp_{T}$) for the signal ($h_{ee}=0.1$) and background
through the process $e^{-}e^{-}\rightarrow e^{-}e^{-}$ at the center
of mass energy $\sqrt{s}=500$ GeV and $\sqrt{s}=3000$ GeV. Here
we take the mass of $H^{--}$ boson equal to the center of mass energy
of the collider. We see from these figures that the signal can be
differentiated from the background by applying the appropriate $p_{T}$
cuts on the electrons.

Since we mainly focus on the resonance production of doubly charged
Higgs boson via the process $e^-e^-\to e^-e^-$ we may neglect the
SM/$H^{--}$ interference terms around the resonance. Because, we
obtained the best limits for the couplings in this case. We see that
after the $p_T$ cuts these terms contribute insignificantly.
Furthermore, in simulation with standard PYTHIA we can't produce
signal and background events together and their interference  at the
same time. Therefore, we will take into account signal and
background separately and make the statistical analysis accordingly.

In Table \ref{table2} and \ref{table3} we show the signal and
background cross sections, for a more quantitative outlook,
depending on the applied $p_{T}$ cuts for $\sqrt{s}=500$ GeV and
$\sqrt{s}=3000$ GeV, respectively. The background cross section
decreases significantly with the higher $p_{T}$ cuts. Some optimized
values can be found by searching for signal to background ratio. We
have used a relation for the determination of optimal $p_{T}$ cuts
$p_{T}>m_{H^{--}}/2-\Delta p_{T}$ where $\Delta p_{T}$ can be
optimized for every mass values to give better signal to background
ratio and resolution as well. With a cut $p_{T}>200$ GeV on the
final state electron we obtain the background cross section
$\sigma_{B}=6.28$ pb and signal cross section $\sigma_{S}=267.0$ pb
for $m_{H^{--}}=\sqrt{s}=500$ GeV and $h_{ee}=0.1$. In the
$e^{-}e^{-}$ collision with $\sqrt{s}=3000$, if we apply a
$p_{T}>1000$ GeV we obtain a smaller cross section for background
$\sigma_{B}=4.94\times10^{-2}$ pb and signal $\sigma_{S}=1.35$ pb
for $m_{H^{--}}=\sqrt{s}$ and $h_{ee}=0.1$.

\begin{table}

\caption{Cross sections for the signal and background with and without ISR+BS
effects depending on the applied $p_{T}$ cuts on final state leptons
at ILC energies.\label{table2}}

\begin{center}\begin{tabular}{|c|c|c|c|c|}
\hline
$p_{T}$(GeV)&
$\sigma_{S}$(pb)&
$\sigma_{S,ISR+BS}$(pb)&
$\sigma_{B}$(pb)&
$\sigma_{B,ISR+BS}$(pb)\tabularnewline
\hline
10&
$2.17\times10^{3}$&
$4.45\times10^{2}$&
$3.00\times10^{3}$&
$2.99\times10^{3}$\tabularnewline
\hline
50&
$2.13\times10^{3}$&
$4.36\times10^{2}$&
$1.27\times10^{2}$&
$1.264\times10^{2}$\tabularnewline
\hline
100&
$1.99\times10^{3}$&
$4.07\times10^{2}$&
$3.41\times10^{1}$&
$3.35\times10^{1}$\tabularnewline
\hline
150&
$1.74\times10^{3}$&
$3.56\times10^{2}$&
$1.52\times10^{1}$&
$1.44\times10^{1}$\tabularnewline
\hline
200&
$1.31\times10^{3}$&
$2.67\times10^{2}$&
$7.29\times10^{0}$&
$6.28\times10^{0}$\tabularnewline
\hline
\end{tabular}\end{center}
\end{table}

\begin{table}

\caption{The signal and background cross sections depending on the applied
$p_{T}$ cuts on final state lepton with and without ISR+BS effects
at CLIC energies.\label{table3}}

\begin{center}\begin{tabular}{|c|c|c|c|c|}
\hline
$p_{T}$(GeV)&
$\sigma_{S}$(pb)&
$\sigma_{S,ISR+BS}$(pb)&
$\sigma_{B}$(pb)&
$\sigma_{B,ISR+BS}$(pb)\tabularnewline
\hline
100&
$6.05\times10^{1}$&
$1.81\times10^{0}$&
$3.23\times10^{1}$&
$3.02\times10^{1}$\tabularnewline
\hline
200&
$5.95\times10^{1}$&
$1.79\times10^{0}$&
$8.46\times10^{0}$&
$6.11\times10^{0}$\tabularnewline
\hline
500&
$5.70\times10^{1}$&
$1.71\times10^{0}$&
$1.44\times10^{0}$&
$4.74\times10^{-1}$\tabularnewline
\hline
1000&
$4.50\times10^{1}$&
$1.35\times10^{0}$&
$3.46\times10^{-1}$&
$4.94\times10^{-2}$\tabularnewline
\hline
\end{tabular}\end{center}
\end{table}

\section{Simulation}

To make our analysis realistic we have smeared the energy of final
state particles by a Gaussian function whose half-width is guided
by the resolution of the electromagnetic (EM) calorimeter
\cite{21}
\begin{equation}
\frac{\delta E}{E}=\frac{10\%}{\sqrt{E}}+0.01\label{eq:9}
\end{equation}
Energy resolution defines the detector's ability to produce narrow
peaks in the spectrum. It is affected by detector geometry and
electronic noise. Resolution is typically measured as the width at
the full width at half maximum for the given energy. In all cases
the smaller the number specified for energy resolution is best and
gives the user a better probability of separating peaks for easier
identification. The invariant mass resolution of the final state
charged leptons can be given by $\delta m_{ll}/m_{ll}\simeq \delta
E/\sqrt{2}E$. Here, we calculate the invariant mass resolution
$\delta m_{ll}=4$ GeV for $m_{ll}=500$ GeV. When reconstructing the
invariant mass of two leptons in the final state we have
incorporated the effects of initial (ISR) and final state radiation
(FSR) using the PYTHIA program \cite{22}. We use the program SMEAR
\cite{23} to smear four-momenta of final state charged leptons,
taking into account detector effects and acceptances. We have
included the detector effects in our calculations for ILC and CLIC
energies. For electrons in the final state, the magnitude of the
output momentum is determined from the smeared calorimeter energy
assuming the mass of the particle is that of an electron. A tracking
device is used for the measurement of four-momenta of the muons in
the final state. The tracking device is modelled in the following
way: (i) the transverse momentum $p_T$ and azimuth angle $\phi$ are
smeared together, taking into account their correlation \cite{24};
(ii) the z-component of the momentum is smeared by projecting the
transverse momentum according to a Gaussian in $1/p_z$. A detailed
discussion of the properties of the detector model implemented in
the present version of SMEAR can be found in \cite{23}. Charged
leptons (electrons and muons) are reconstructed within the
acceptance range in pseudo-rapidity $|\eta|<2$. Momentum resolution
of the tracker is given depending on the scattering angle, and
assumed as 0.02\% for $\theta>40$ degrees. The efficiencies for
charged particles are assumed to be 98\%.

\begin{figure}
\begin{center}\includegraphics[
  scale=0.8]{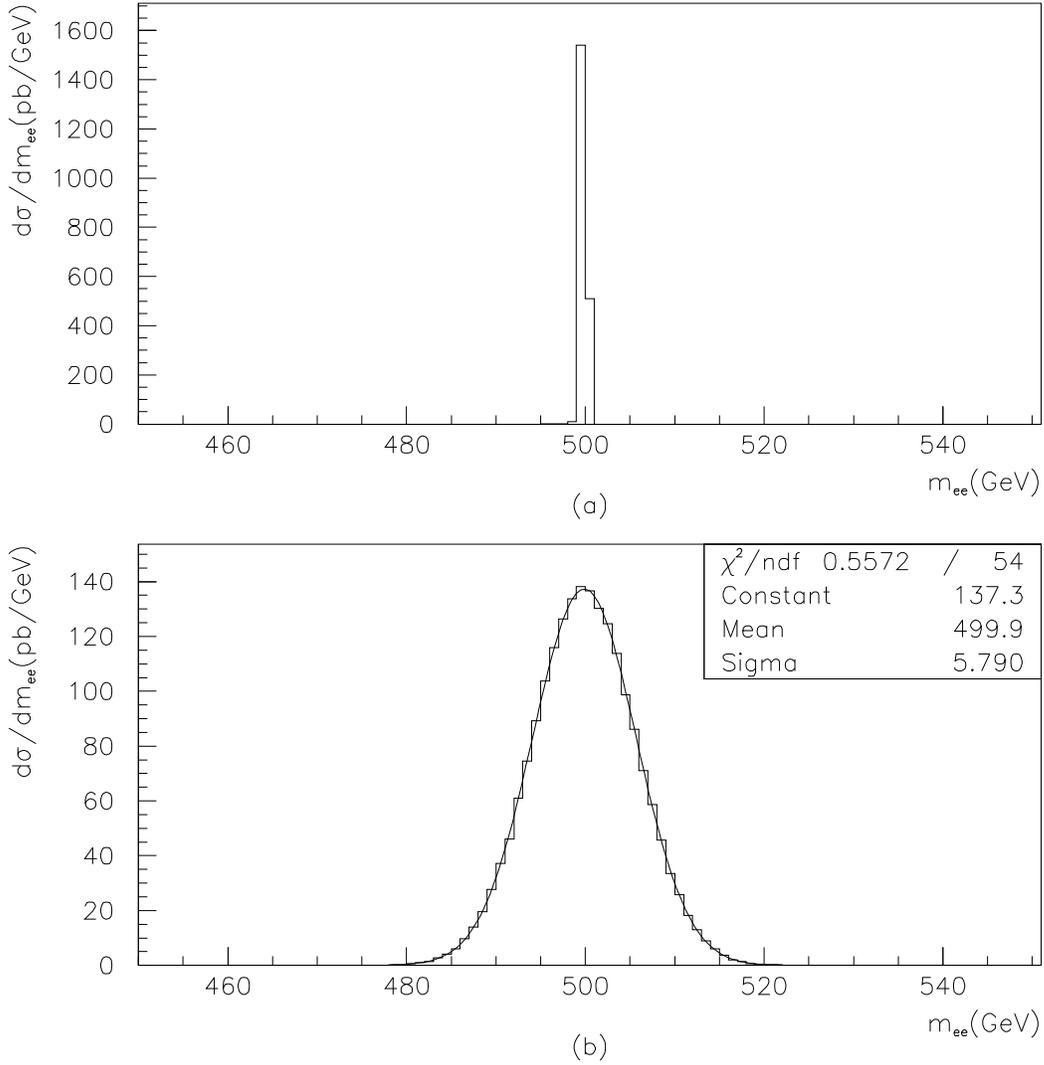}\end{center}
\caption{Invariant mass distribution of two electrons at $\sqrt{s}=500$ GeV
for $h_{ee}=0.1$: (a) signal as a narrow peak, (b) signal with detector
effects and a Gaussian fit. \label{fig6}}
\end{figure}

We show the invariant mass distribution of the $e^-e^-$ pair for the
above choice of parameters in Fig. \ref{fig6}. Here the distribution
peaks correspond to the mass values of doubly charged Higgs bosons.
Invariant mass of two electrons is smeared according to the detector
effects to be observable experimentally.

\begin{figure}
\begin{center}\includegraphics[
  scale=0.8]{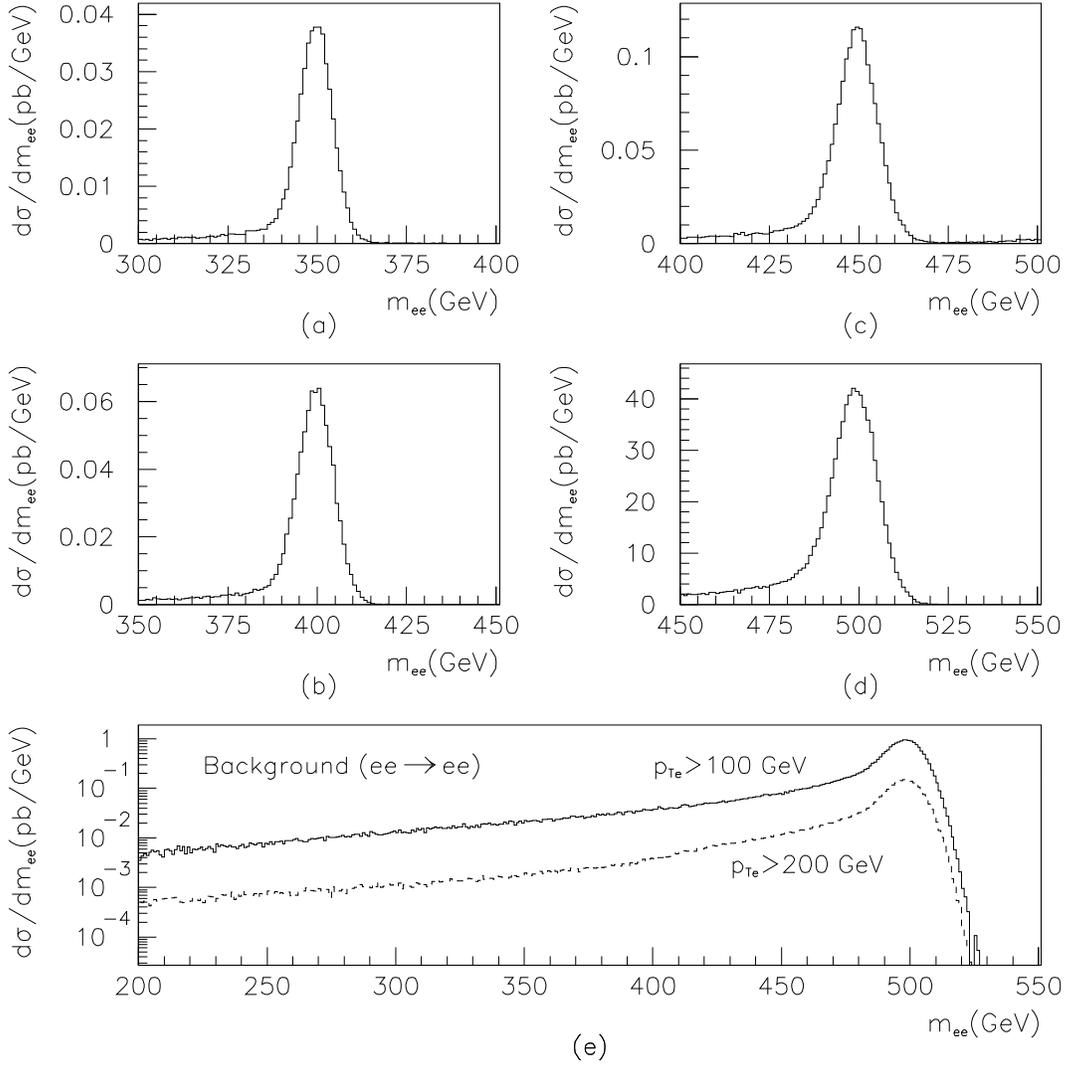}\end{center}

\caption{The invariant mass distribution of two electrons in the
final state at the ILC energy $\sqrt{s}=500$ GeV; (a-d) the upper
four plots present the signal with $m_{H^{--}}=$350, 400, 450, 500
GeV. In (d) we show the signal distribution for the coupling
$h_{ee}=0.1$. (e) the lower plot is for the background
($ee\rightarrow ee$) with $p_{Te}>100$ GeV and $p_{Te}>200$ GeV.
\label{fig7}}
\end{figure}

\begin{figure}
\begin{center}\includegraphics[
  scale=0.8]{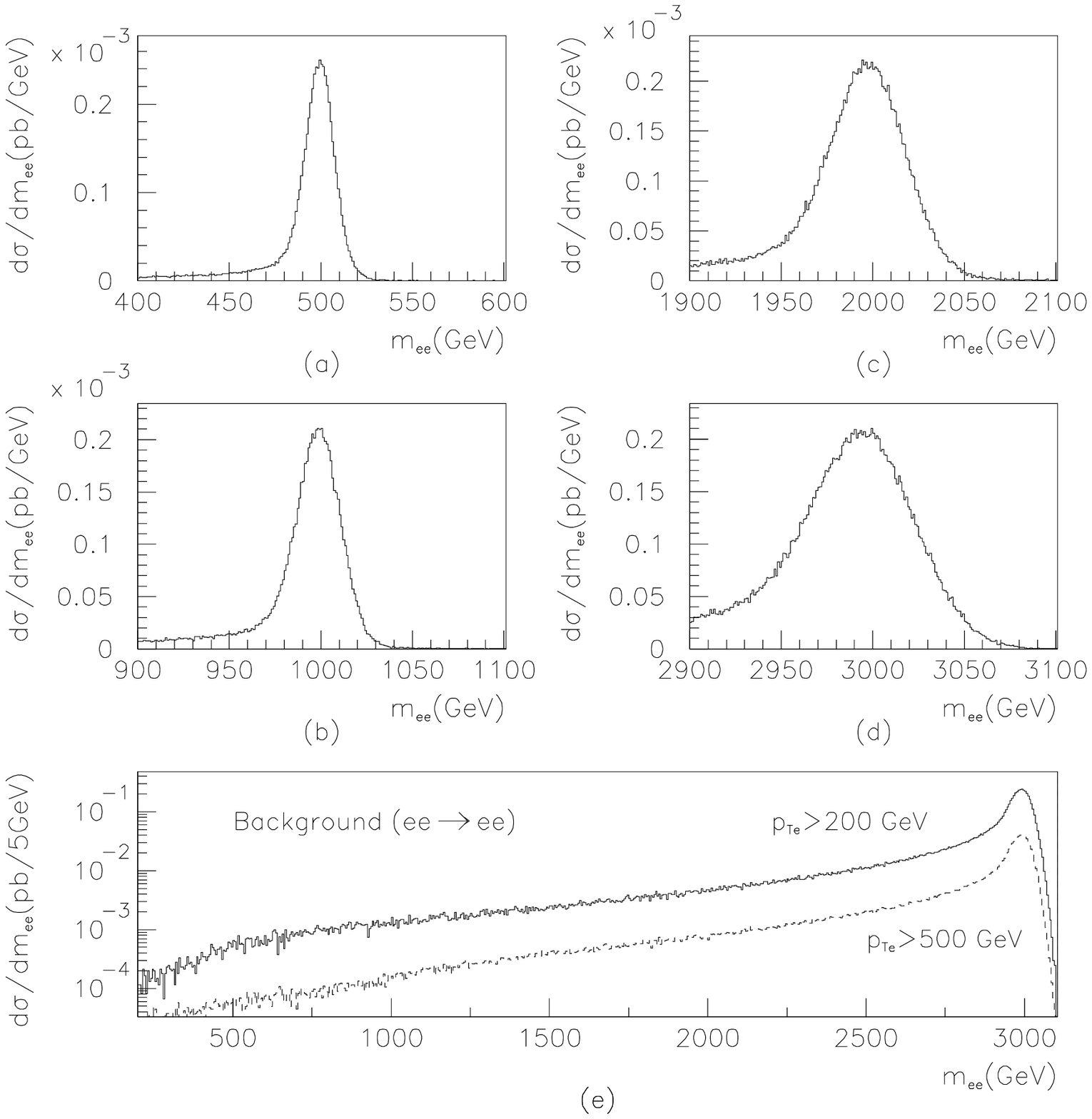}\end{center}
\caption{The invariant mass distribution of two electrons in the
final state at the CLIC energy $\sqrt{s}=3000$ GeV; (a-d) the upper
four plots present the signal with $m_{H^{--}}=$500, 1000, 2000,
3000 GeV. In (d) we show the signal distribution for the coupling
$h_{ee}=0.1$. (e) the lower plot is for the background
($ee\rightarrow ee$) with $p_{Te}>200$ GeV and $p_{Te}>500$ GeV.
\label{fig8}}
\end{figure}

The invariant mass distribution of two electrons in the final
state at the ILC and CLIC energies are shown in Fig. \ref{fig7}
and \ref{fig8}. Since we include the ISR+BS and smearing effects
to the production cross sections there appears a coupling
dependence moderately at very resonance point. In Fig. \ref{fig7},
we present the signal with $m_{H^{--}}=$350, 400, 450, 500 GeV and
the background ($e^-e^-\rightarrow e^-e^-$) with $p_{Te}>100$ GeV
and $p_{Te}>200$ GeV. In Fig. \ref{fig8}, we present the signal
with $m_{H^{--}}=$500, 1000, 2000, 3000 GeV and the background
with $p_{Te}>200$ GeV and $p_{Te}>500$ GeV. We see from these
figures that the signal peaks over the background at resonance
even for smaller couplings. Applying appropriate $p_{T}$ cuts on
the electron from the background process ($e^-e^-\rightarrow
e^-e^-$) we can also detect the signal for smaller mass values of
the doubly charged Higgs boson.

\begin{figure}
\begin{center}\includegraphics{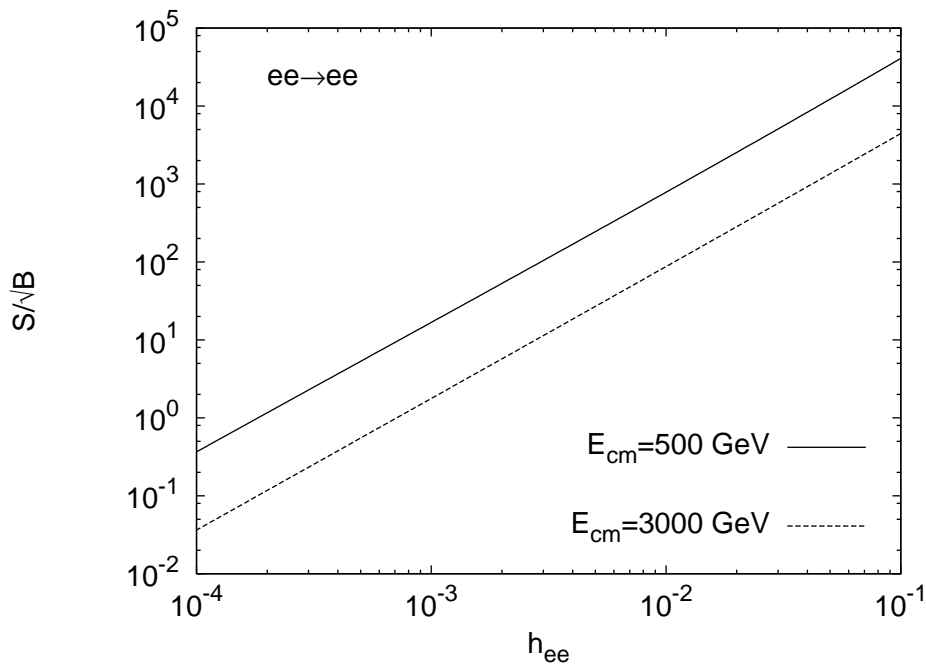}\end{center}

\caption{Signal significances depending on the coupling $h_{ee}$ for
doubly charged Higgs boson (when produced at resonance) at the
collider energies $E_{cm}=500$ GeV and $3000$ GeV.\label{fig9}}
\end{figure}

Since we simulate signal and background events we define the signal
significance as $SS=N_{S}/\sqrt{N_{B}}$ where $N_{S}$ and $N_{B}$
are the signal and background events in the chosen mass bin,
respectively. Taking the corresponding integrated luminosities for
ILC and CLIC $e^-e^-$ options $L_{int}=10^{4}$ pb$^{-1}$ and taking
into account the detection efficiencies for the leptons we calculate
the number of events for the signal and background leading to the
significances given in Table \ref{table4}. Here, we find $\delta
m_{ll}>>\Gamma_{H^{--}}$, and therefore the signal significance is
affected by the ISR+BS, FSR and smearing. If we consider lepton
flavour conservation we can calculate the signal significance
depending on the mass and coupling $h_{ee}$ of the doubly charged
Higgs boson by applying appropriate $p_{T}$ cuts. We may define the
bins of width $\Delta m$ around the signal peak in the invariant
mass distributions for each value of $m_{H^{--}}$. We then look at
the fluctuations in the SM background in that bin and compare the
corresponding rate for the signal. This procedure can be repeated
for different couplings to find limiting values. The reaches of the
coupling constant for the mass values of doubly charged Higgs boson
smaller than the center of mass energies are given in Table
\ref{table4}. The resonance point worth a detailed study since
smearing, initial and final state radiation affect the coupling
dependence of the cross section. We estimate the reach of the
coupling at which the resonances in the invariant mass distribution
can be identified over the fluctuations in the SM background at
$3\sigma$ level. We see from Fig. \ref{fig9} that we can probe the
doubly charged Higgs boson (when produced at resonance) coupling
down to $h_{ee}=3\times10^{-4}$ and $h_{ee}=10^{-3}$ with 95\% C.L.
at the center of mass energy $E_{cm}=500$ GeV and $E_{cm}=3000$ GeV,
respectively.

\begin{table}

\caption{Signal significances [$SS(h_{ee})=S(h_{ee})/\sqrt{B}$] for
the process $e^{-}e^{-}\rightarrow e^{-}e^{-}$ at the center of mass
energy $\sqrt{s}=500$ GeV and $\sqrt{s}=3000$ GeV taking the
integrated luminosity of $L_{int}=10^{4}$pb$^{-1}$. \label{table4}}

\begin{center}\begin{tabular}{|c|c|c|c|c|c|}
\hline
\multicolumn{3}{|c|}{$\sqrt{s}=500$ GeV, $p_{Te}>$100 GeV}&
\multicolumn{3}{c|}{$\sqrt{s}=3000$ GeV, $p_{Te}>$200 GeV}\tabularnewline
\hline
$m_{H^{--}}$(GeV)&
$\Delta m$(GeV)&
$SS(0.05)$&
$m_{H^{--}}$(GeV)&
$\Delta m$(GeV)&
$SS(0.1)$\tabularnewline
\hline
350&
10&
4.5&
500&
18&
6.3\tabularnewline
\hline
400&
11&
4.8&
1000&
26&
5.4\tabularnewline
\hline
450&
12&
6.2&
2000&
44&
3.9\tabularnewline
\hline
\end{tabular}\end{center}
\end{table}

\begin{figure}
\begin{center}\includegraphics[%
  width=150mm,
  height=120mm]{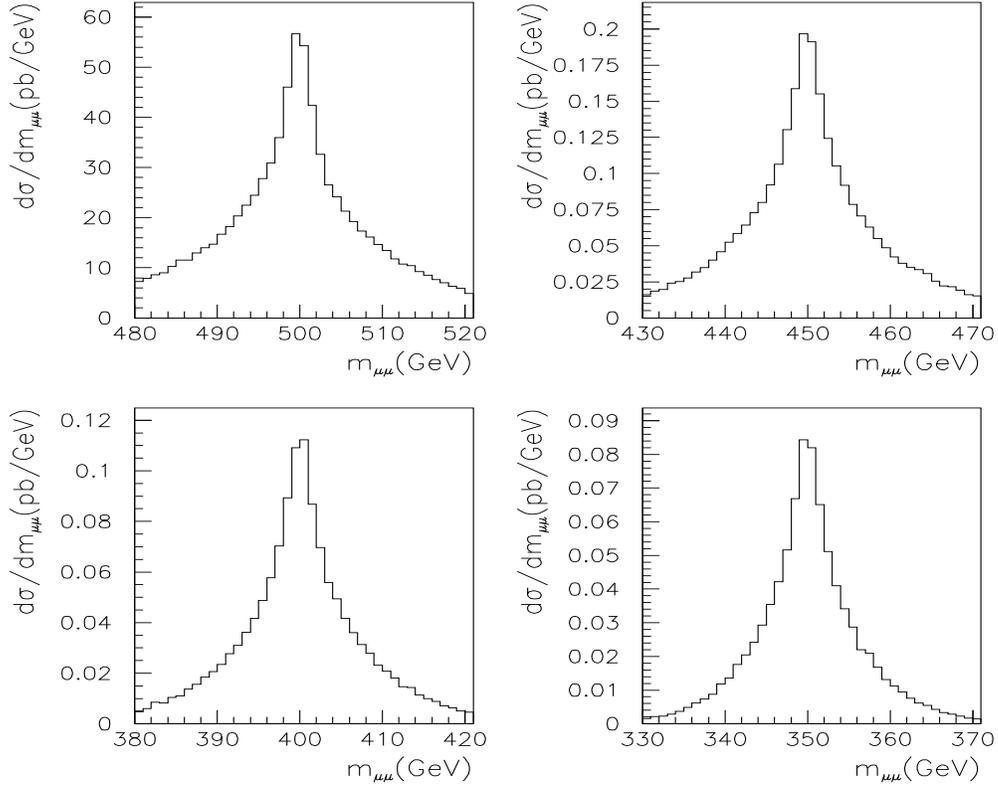}\end{center}

\caption{Invariant mass distribution of the two-muon in the final state for
different mass values of doubly charged Higgs boson at the center
of mass energy $\sqrt{s}=500$ GeV. \label{fig10}}
\end{figure}

\begin{figure}
\begin{center}\includegraphics[
  width=14cm,
  height=10cm]{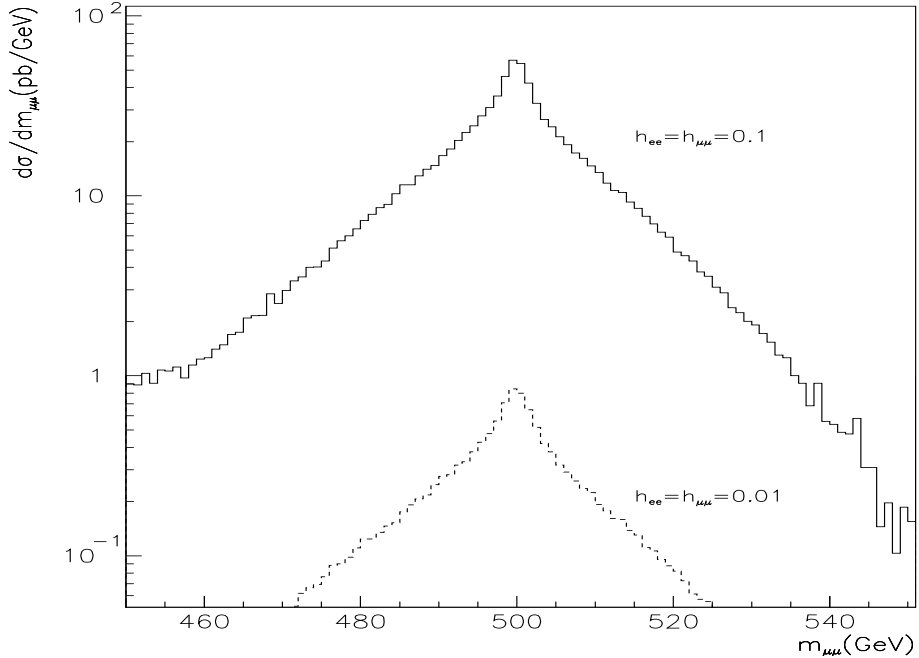}\end{center}

\caption{Invariant mass distribution of two-muon around the
resonance. The curves are obtained for different couplings $h_{ll}$
by taking into account ISR+FSR and smearing effects at
$\sqrt{s}=500$ GeV.\label{fig11}}
\end{figure}

\begin{figure}
\begin{center}\includegraphics{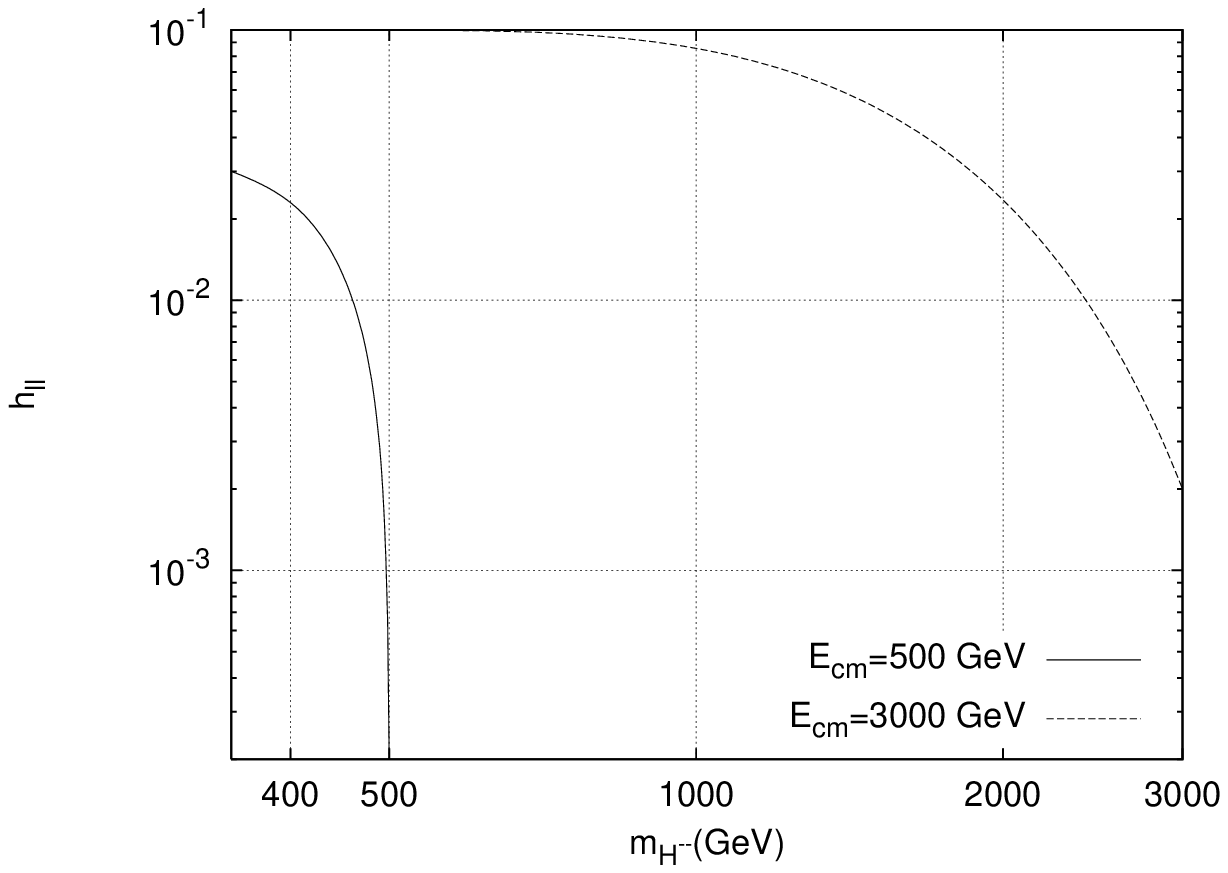}\end{center}

\caption{Contour plot of the doubly charged Higgs boson coupling $h_{ll}$
and mass $m_{H^{--}}$ for the lepton flavor violating process $e^{-}e^{-}\rightarrow\mu^{-}\mu^{-}$
at ILC and CLIC energies.\label{fig12}}
\end{figure}

For the process $e^{-}e^{-}\rightarrow\mu^{-}\mu^{-}$ the
differential cross sections can be obtained as shown in Fig.
\ref{fig10}. As can be seen from this figure two-muon invariant mass
distribution shows a sharper peak than that of two-electron. In Fig.
\ref{fig11} invariant mass distributions have been shown for
different couplings at the resonance $\sqrt{s}=500$ GeV. In case of
the lepton number violating process
$e^{-}e^{-}\rightarrow\mu^{-}\mu^{-}$ there is no significant SM
background. We obtain the 95\% C.L. contour by taking Poisson
variable as the observed events with the mean of 9. In Fig.
\ref{fig12} the discovery contour of $h_{ll}$ and $m_{H^{--}}$ are
plotted for the lepton number violating process including ISR+FSR
and smearing effects. As can be seen from Fig. \ref{fig12} the
accessible region of the contour plot can be enriched for higher
integrated luminosities at future linear collider options.

If we use an electromagnetic calorimeter with a larger energy
resolution, we obtain a smaller cross section for the signal and a
larger one for the background ($e^-e^-$) to satisfy 95\% CL.
Therefore, our results for $S/\sqrt{B}$ will be changed
accordingly and discovery region for doubly charged Higgs boson
will become more restricted.

\section{Conclusions}

We have investigated the potential of the $e^{-}e^{-}$ running
mode of the linear colliders for discovering and studying doubly
charged Higgs bosons. In $e^{-}e^{-}$ collisions the beams will
deflect each other leading to a reduction in luminosity comparing
to the $e^{+}e^{-}$ collisions in which two beams focus each
other. For the part of the luminosity spectrum close to the
nominal center-of-mass energy, the reduction in $e^{-}e^{-}$ mode
is thus a factor of about 2.5 compared with $e^{+}e^{-}$ \cite{2}.
The particularly clean environment of the $e^{-}e^{-}$ collisions
justifies here the neglect of systematic errors. For luminosity
higher than what we have used in our calculation the reach for
couplings can be further enhanced.

Using highly polarized electrons in the initial state make the
resonances for doubly charged Higgs bosons observable down to quite
smaller couplings. It has the benefit that at the same time one may
also switch off most of the SM background processes (including
$W^{-}$ bosons) which could mask the discovery of new particles. The
linear collider sensitivity to doubly charged Higgs boson properties
is best when running at the resonance $\sqrt{s}=m_{H^{--}}$. At the
ILC and CLIC energies the couplings to charged leptons can be tested
down to $3\times 10^{-4}$ and $10^{-3}$ at the resonance,
respectively. Our analysis depends on fixed center of mass energy of
the machine. Around the resonance there is still radiative return to
the doubly charged Higgs boson resonance due to the initial state
radiation, which makes it possible to discover these particles for
small couplings (down to $5\times 10^{-2}$ for $m_{H^{--}}=1500$ GeV
at $\sqrt{s}=3000$ GeV) without having a collider energy scan. The
numerical analysis we performed is also relevant for any model which
contains a Higgs triplet (e.g. Higgs triplet model, L-R symmetric
models and little Higgs models).

It is apparent that if a doubly charged Higgs boson is found at
the LHC, an $e^-e^-$ collider would separately and in combination
provide enormously important information concerning the structure
and interactions of the Higgs sector. This also illustrates an
important complementarity between the ILC or CLIC and the LHC. The
discovery of a $H^{--}$ prior to the construction and operation of
the $e^+e^-$/$e^-e^-$ collider would be very important in
determining the range over which good luminosity and good energy
resolution for $e^-e^-$ collisions should be a priority. Our
calculations show that the signal significance $SS~\sqrt{L}$ can
be rescaled for a lower luminosity in case of a limited time of
e-e- running mode of ILC and CLIC.

On discovering a $H^{--}$ it would be important to measure the
absolute value of $h_{ij}$, therefore a detailed experimental
simulation of the decay modes is encouraged at both hadron and
lepton colliders.

\begin{acknowledgments}
This work is supported in part by the Turkish State Planning Organization
(DPT) under the grant No DPT-2006K-120470.
\end{acknowledgments}

\end{document}